# Reorienting Learning Game Design in Design-Based Research : a Case Study

Nadine Mandran [a*][0000-0002-8660-3827], Estelle Prior [b][0000−0002−2895−2690], Eric Sanchez [b][0000-0002-3819-6681], Mathieu Vermeulen [c][0000−0003−3646−1741]

[a] LIG, University of Grenoble Alpes, Saint Martin d'Hères, France, [0000-0002-8660-3827]

[b] LIP/TECFA, University of Geneva, LIP, Carouge, Switzerland

[c] IMT Nord Europe, PIRP, University of Lille, Lille, France

Corresponding author: Nadine Mandran , University of Grenoble Alpes, Saint Martin d'Hères, France
nadine.mandran@univ-grenoble-alpes.fr

# A case study to reorient LG design in Design Based-Research

**Abstract.** One of the main difficulties remains the collaboration between the various experts involved in designing the Learning Games (LG). Our literature review focuses on the pitfalls and principles that have been identified by various authors in learning games design. Based on this review, a prototype was designed to support the LG design process and to study more precisely the collaboration between actors (teachers, researchers, game designers, data analyst and computer scientist). Indeed, according to the state of the art, the skills and knowledge involved in design are difficult to integrate. It has been tested in a real-world scenario for designing learning games to teach algorithmic. Through participant observation in thirty-three workshops involving nine experts, we were able to identify recurring pitfalls as we applied the recommendations in the literature. The analysis of these workshops led to propose eight principles aimed at facilitating the collaboration between the learning games design process and re-evaluating research on its.

**Introduction**

Learning Games (LG) defined as "a virtual environment and a gaming experience in which the contents that we want to teach can be naturally embedded with some contextual relevance in terms of the game-playing" (Fabricatore, 2000) are becoming increasingly prevalent in teaching. While LG are increasingly present in training curricula, their development remains complex due to the involvement of diverse and sometimes very distant expertise (Vosinakis et al., 2020).

Furthermore, many researches in Technology Enhanced Learning (TEL) or in educational sciences are interested in LG as a research object. The Design-Based Research (DBR) used in this research context (M.-C. Li & Tsai, 2013; Mandran N. et al., 2022; The Design-Based Research Collective, 2003; F. Wang & Hannafin, 2005) proposes involving teachers in the design of LG, development of the research problem, and analysis of data. The use of DBR raises the question of how to include the research approach during the design of LG with others expertises.

LG design can be based on various generic methods (e.g. meta-design, user-centered design) and some others specific of TEL (e.g. Paquette et al., 1999, Marne et al., 2012, Marfisi-Schottman et al., 2010, Vermeulen M. et al., 2018). Most often reused by their authors, we can arise the question of the reuse by others in a DBR research context and more precisely on the collaboration between experts.

First, we present a literature review of pitfalls and recommendations to design LG. Then, we describe a prototype designed and used to conduct our study. Next, we present our research method and the field experiment. Finally, based on literature and our observations, we propose height principles to guide the design of LG in DBR context that involve researchers and practitioners.

**Literature review**

In literature review, we present a list of problems and recommendations identified for the design of LG and more specifically on collaboration between specialists.

We have identified six categories of problems mainly related to the collaboration:

(i) The design team integrates multiple specialists with different disciplines, expectations, experiences, interests, goals and priorities (Arnab et al., 2012; Fisher, 2019; Jaccard, Suppan, & Bielser, 2021; Jaccard, Suppan, Sanchez E., et al., 2021; Ke et al., 2019; Klerks et al., 2022; Morard

& Sanchez E., 2021; Tahir & Wang, 2020; Vosinakis et al., 2020; Wake et al., 2018). Certain skills may also be lacking to drive the process (Fisher, 2019).

(ii) This interdisciplinarity approach can result in communication issues within the team. The language used, such as vocabulary and terminology, regarding the process may be specific to each community (Abdelali et al., 2016; Jaccard, Suppan, Sanchez E., et al., 2021; Ke et al., 2019; Morard & Sanchez E., 2021). These specialists may hold different perspectives on the process (Muñoz et al., 2022), or adopt epistemological positions that offer different perspectives (Abrahamson & Chase, 2015; Ke et al., 2019). Also, authors have noticed the lack of models or methods for involving the users in the process (Kim et al., 2021; Yang et al., 2021).

(iii) Integrating users such as teachers and students, and keeping them in the game design team can be challenging (Alenljung & Maurin Söderholm, 2015; Kalmpourtzis, 2019; Klerks et al., 2022; Muñoz et al., 2022; Sim et al., 2015).

(iv) The expertise of specialists is limited to their expertise fields, which can result in communication difficulties (Tahir & Wang, 2020). For instance, teachers may not possess sufficient experience, expertise or training in game design or implementation (Baradaran Rahimi & Kim, 2019; Earp et al., 2016; Fu et al., 2022; Karoui et al., 2020; Kim et al., 2021; Q. Li, 2018; Pombo & Marques, 2021; Wake et al., 2018). On the other hand, designers may lack pedagogical knowledge (Barreto, 2015; Fisher, 2019; Sillaots & Maadvere, 2013).

(v) Balancing and coherently integrating playful and educational elements can be a challenge (Abdelali et al., 2016; Barreto, 2015; Jaccard, Suppan, Sanchez E., et al., 2021; Q. Li, 2018; Pirker & Gütl , 2015; Wake et al., 2018). The challenge is to maintain consistency between "the skill or content to be learned" and "the elements of fantasy and play" (Baradaran Rahimi & Kim, 2019;

Ke et al., 2019). In the game, this knowledge can be represented through an implicit metaphor linked to a universe relevant to the knowledge learned (Bonnat et al., 2023).

(vi) Research on LG is seldom initiated in collaboration with field actors. This is because little research considers their needs (Sanchez E. et al., 2017) or the realities of the field (Ke et al., 2019).

To overcome the aforementioned obstacles, these authors provide recommendations that we have chosen to grouped into six categories:

(i) To begin the process, the initial problem should be define and an objective should be identified. It is recommended to consult with specialists inside or outside the team and end users, such as teachers, to ensure its feasibility (Plank et al., 2011; Van Dooren et al., 2016). Additionally, one of the specialists should ensure that the project' scope is well understood (Kucirkova, 2017).

(ii). To design a LG, it is necessary to articulate specialized knowledge. It is not a question of adding expertise and knowledge one after the other or of dividing the work but of coordinating the skills and knowledge of specialists (Ke et al., 2019). It is recommended to adopt a common vocabulary (Alenljung & Maurin Söderholm, 2015; Jaccard, Suppan, & Bielser, 2021; Ke et al., 2019), and to share the information necessary to carry out the design process (Fu et al. al., 2022; Jaccard, Suppan, Sanchez E., et al., 2021; Karoui et al., 2020; Ke et al., 2019; Morard & Sanchez E., 2021; Muñoz et al., 2022; Pombo & Marques, 2021; W.-L. Wang et al., 2010).

(iii) It is necessary to ensure that the team adopts a "shared design culture" throughout the process (Ke et al., 2019, p. 17). To achieve this, specialists' perspectives (e.g. priorities, intentions, agenda) need to be adjusted to the process. Workshops dedicated to collaborative design help to ensure this adjustment (Klerks et al., 2022). If partners from outside the project are involved, it is necessary to ensure that their objectives are also adjusted to those of the project (Muñoz et al., 2022). It is important to specify and select the models and concepts to be mobilized (Jaccard, Suppan, Sanchez E., et al., 2021; Ke et al., 2019)..

(iv) To support the design and development of LG, roles should be adopted, negotiated and defined without becoming too rigid (Baradaran Rahimi & Kim, 2019; Jaccard, Suppan, & Bielser, 2021; Laakso et al., 2021). A specialist in the method to be followed or the tool to be used should be included to guide the team through the process (Goudswaard et al., 2019; Jaccard, Suppan, & Bielser, 2021).

(v) To effectively understand the process, it is recommended to (a) plan tasks need; (b) allocate them among specialists needs in a fluid and spontaneous manner; (c) ensure that everyone is aware of their tasks and those of others; (d) identify the missing skills in the team (Jaccard, Suppan, & Bielser, 2021; Jaccard, Suppan, Sanchez E., et al., 2021; Wake et al., 2018). Work can be structured in sub-teams (Koutsabasis et al., 2022).

(vi) To support the process is necessary to have some specific instruments. Negotiation and decision-making can be supported through co-design workshops (Klerks et al., 2022), brainstorming sessions (Ke et al., 2019), voting system (Jaccard, Suppan, Sanchez E., et al., 2021) or tools for identifying trade-offs (e.g. LEAGUE Thinking Maps) (Tahir & Wang, 2020). Specialists should be able to record, analyze, share and archive the explorations and experiments carried out during the problem-solving process (Ke et al., 2019). Baradaran Rahimi and Kim (2019) recommend using a design journal to trace the design.

Although the literature review identifies pitfalls and recommendations, we are concerned with the following research questions: - What are the other limiting factors in LG design that are not yet documented in the literature? - Should we reconsider LG's collaborative design in light of existing models tested in the field, and if so, according to what design principles? These two questions arise in a context where the game design serves a teaching situation and a research situation. To do this, we built a prototype based on recommendations and design models proposed by research teams. It is also a way to test the reusability of these models.

**Material: Co.LAB prototype**

The Co.LAB prototype designed to support our study include a set of recommendations from previous work. It is based on several design methods including "6 facets" (Marne, 2014); the "7 steps" (Marfisi-Schottman et al., 2010); « DISC » (Vermeulen M. et al., 2018); « 10 steps » (Plumettaz-Sieber, Hulaas, et al., 2019); « co.LAB model » (Jaccard, Suppan, & Bielser, 2021) ; "THEDRE" (Mandran N., 2018). All the methods identified propose steps to structure the design process. They indicate which specialists to mobilize and propose tools to support the work. Nevertheless, working methods, constraints and skills of each specialist are rarely described. Similarly, there is no description of the negotiation and decision-making processes. While the model of Plumettaz-Sieber, Hulaas, et al. (2019) lists teachers as partners in working on the research there is no tools to support this work. Furthermore, none of these models propose steps or tasks for conducting research based on field questions posed by teachers.

So, the prototype includes a set of tasks categorized into 7 subprocesses (Fig1): 1) Starting the project; 2) Describing the didactic and pedagogical components; 3) Designing the game; 4) Describing the context of the game; 5) Writing the rules of the game; 6) Conducting research; 7) Monitoring the project.

Each sub-process consists of a series of tasks, the order of execution is not strictly linear or iterative but merely systemic. Instead, it involves moving back and forth between tasks. For instance, the definition of learning outcomes may need to be revised multiple times after being initially specified. However, some tasks rely on others. For example, "Define learning outcomes" depends "Specify the context". As a result, the design of the game is broken down into 5 subprocesses: performing ideation; describing the game's metaphor; detailing the nature of the activities; describing the missions and levels; and testing the game.

Each sub-process makes available a set of documents that are shared among team members. They include a series of questions that guide activity, and ensure process traceability and decisions making. These instruments intended to support reflection are integrated into digital platforms (e.g. Google drive, Miro) that allow team members to work in person or remotely, synchronously or asynchronously. Some instruments were designed ad hoc for our experiment. Also some tools are included during the field experiments; they are proposed by some members of experiment.

As for the people involved, we have identified several roles from the aforementioned work: the teacher who will integrate the game into his lessons and who knows the context; the didactic expert who knows learning models and methods; the game designer; the game developer; the researcher. In practice, these roles overlap. For example, teachers can specify the context of the game and have knowledge of game design. Likewise, a researcher may have expertise on the content to be taught in addition to their research skills. We have also identified the role of the project manager, who is responsible for steering the project and planning activities.

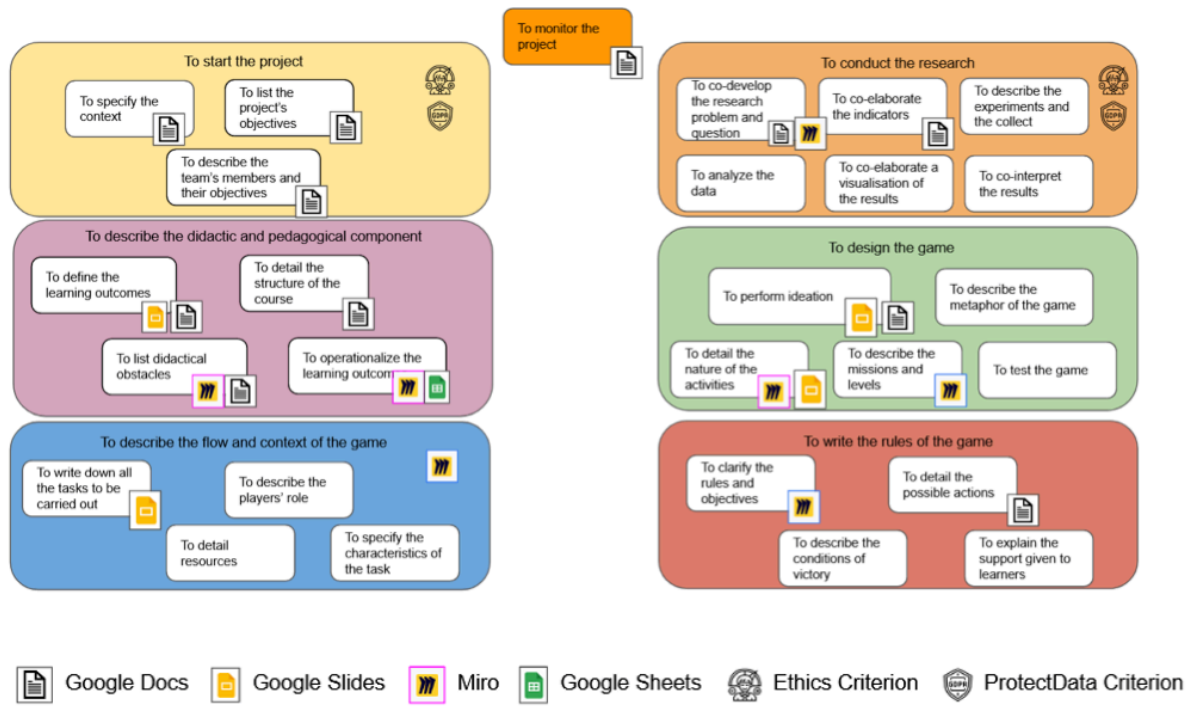

Figure 1 - The 7 subprocesses of game co-design proposed on the co.LAB model

**Method of conducting research**

Our research aims to identify the pitfalls encountered during the design of LG within collaborative research, such as DBR. Our epistemological posture is the pragmatic constructivism (Avenier & Thomas, 2015). It is based on the consideration of reality and the human experience within this reality. The aim of scientific construction is "to develop intelligible models of human experience, offering suitable and viable reference points". We have chosen this posture because the observation of the interactions between actors is central to identify the difficulties and the factors limiting the design process.

*Study objectives*

In order to answer our research questions, we set three objectives for our study:

- Testing the implementation of existing recommendations and models for the design of LG in a DBR context, through co.LAB prototype

- - Identifying collaboration difficulties between specialists

- Identifying unresolved points that deserve to be studied more deeply.

We specify that the objective of our study is not to improve the prototype, it is a tool to highlight the difficulties. However, we were able to make some changes so as not to get stuck. These changes have been traced.

*Method of data production and analysis*

As, our study takes place in a real context, we used a method of data production based on participant observation (Copans, 2012). In this method, the researchers not only observe but also participate in the study. In our case, researchers participate in the different stages of the LG design.

The data produced are qualitative data. They include 1) video recordings of each workshop, 2) semi-structured interviews and 3) answers to open-ended questions in a questionnaire. Semi-structured interviews were carried out at different stage. The purpose of them was to study the perception of the work and the difficulties encountered, and to gather suggestions for improvements. We interviewed three participants: a person experienced in LG design who had been questioned about his development skills, a person new to the design of LG who was taking part in the process as a researcher in computer science didactics and a computer science teacher interested in using the game in his course.

When version one of the LG was completed, an open-ended questionnaire was e-mailed to all participants to gather their opinions on the work and the difficulties that had blocked the process. A total of nine people responded to this questionnaire. This questionnaire should be seen as a qualitative method because the questions are only open-ended and the number of people questioned corresponds only to the project participants.

To analyze data, we use an inductive reasoning method. Induction consists of drawing conjectures from observation that need to be discussed. This reasoning is based on the observation of facts and the identification of phenomena, and seeks explanations for the phenomena observed in the field. These analyses do not require a theoretical model chosen *a priori*. The video, the interviews and open-ended question are analysed in this way.

*Participants*

The participants include: computer science teachers (T1, T2); a head of computer science courses (RES1); a researcher in computer science didactics (RD1); two researchers in LG design and game-based learning (RJ1, RJ2); an expert in research methods (RM1) and; a game designer (GD1). Two doctoral students (P1; P2) doing their thesis on the collaborative design process of LG are in charge of preparing and following up the meetings. Other experts were brought in on an ad hoc basis, during the workshops. We purposely chose participants who were not used to working together, as this helped to avoid any implicit biases that could have potentially distorted our study. Some members had never participated in the design of a LG before.

The ethics commission of the UXanonymous has approved our study.

*Conduct of the study*

Our study takes place to design a LG for teaching undergraduate students the basics of JAVA programming. To study this design process and the difficulty we used the prototype. Thirty-three workshops were organized, and the project is still ongoing (table 1 in the annex). The workshops varied in duration from 45 minutes to 3.5 hours, and were conducted remotely via video-conferencing. A first version of the game was tested in class after workshop 26.

The study was carried out in three stages:

1) Before each workshop, the PhD students ("facilitators") drew up a document setting out the objectives of the workshops, the working method to be used (e.g. presentation, brainstorming, etc.) and the documents required for the exchanges. These tools were either documents associated with the initial prototype or other documents created to meet an unidentified need.

2) At the beginning of the workshop, the previous progress of work is presented to the participants. The objectives of the workshop and documents were presented to the participants. Thus, participants to familiarize themselves with the documents and fill them in collaboratively. Then the discussions and activities began

3) After the workshop, some improvements were also made to the design method, the instruments and the prototype. If any issues arise during the workshop, the facilitators contacted the participants to ensure continued engagement.

Figure 2 shows the sequence of workshops held. Each block corresponds to a workshop. The color code corresponds to the main issue raised. In the case of development meetings (shown in gray), the colored rectangles indicate the points discussed. Two symbols indicate when the prototype was modified and the game developed or tested.

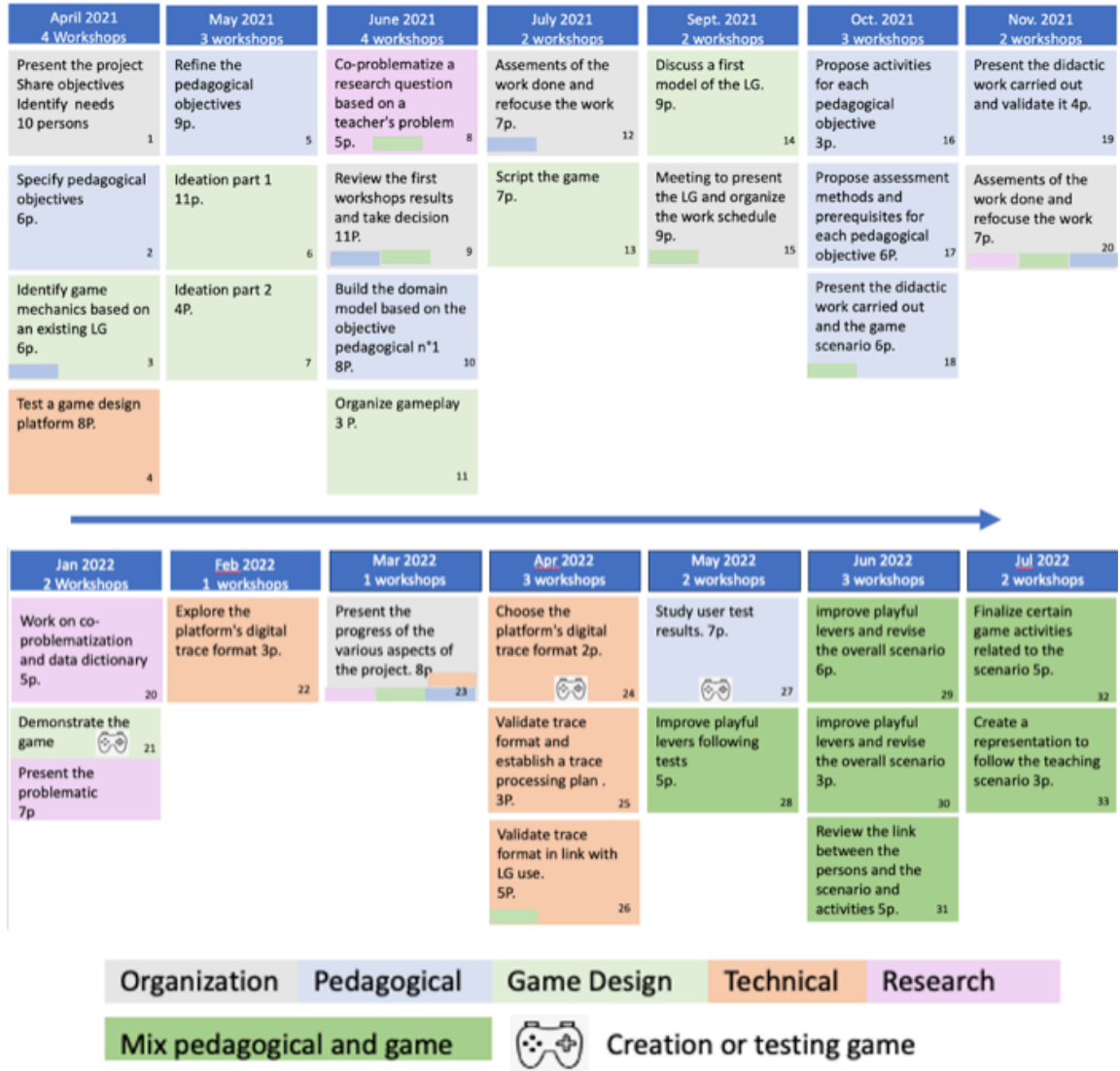

Figure 2 - Sequence of workshops, objectives and number of participants

**Results**

To analyze our purely qualitative data, we proceed inductively without an a priori model to reveal difficulties not known in the literature. Choosing an a priori analysis model would bias our results. We identify themes and subthemes, which are ideas rather than words. It is about gradually answering the question "what is fundamental in this statement, in this text, what is it about" (Paillé

& Mucchielli, 2016). We then write a summary based on the identified themes and sub-themes, and we illustrate them with verbatims. The results focus on the difficulties of collaboration between participants as this is a fundamental principle of DBR. The results are presented in 4 categories of difficulties linked to: 1) Institutional context, technical context and participants, 2) Sharing the design process, leader and instruments, 3) Conduct of research and data, 4) Unshared knowledge

*Institutional context, technical context and participants*

Concerning the institutional context, teachers were faced with difficulties with their management regarding the revision of programs and institutional mergers. At first, discussions about these difficulties took priority over discussions about game design.

The initial technical choices made for the game proved to be blocking (e.g. digital traces could not be collected in a suitable format). The team tested different development environments before choosing a satisfactory one. "*I found that there were a lot of questions about technical choices without really understanding everything....If these elements are constraints, they must be known from the start of the design, otherwise the scripting risks turning out to be completely unsuitable*" (S3). "*At the beginning: find the most suitable software. Next: find out how to recover the necessary traces*" (S7).

Although the project was not funded, the participants remained participated in numerous workshops. However, they reported that they did not "*find their place*" (S4). Their roles and responsibilities were not clear enough, and the heterogeneity of the profiles led to misunderstandings and on concepts and priorities. In addition, adapting to a common working method was challenging. After several sessions, it became evident that there was a lack of game development skills among the team members. Furthermore, decision-making was one of the difficulties as none of the participants felt legitimate to make a decision. "*The variety of habits and working methods of participants from different disciplines required everyone to adapt to common functioning*" (S7).

Furthermore, it was found that the role of the game designer is poorly defined. This role can combine technical, artistic, and playful skills, making it challenging for a single designer to possess all of them. Thus, if the game designer lacks certain technical skills, such as coding, it is advisable to involve a developer. "*The question of skills (of the game designer) was not considered sufficiently during the first stages. I feel like the development environment that will be used should be considered sooner*" (S6).

***Sharing the design process, leader and instruments***

During the first workshop, we presented the entire process to the participants, along with the tool to conduct it (Fig1), the associated tasks and the relevant instruments. However, some participants requested further clarification on the provided process such as connecting the learning outcomes to the game metaphor. "*I knew the theoretical aspects of a collaborative game design process but I didn't know the practical aspects (it was my first game design)*" (S1). Several of us were in this case of first game design, maybe there too it was a brake. Participants stated that they had "*a lack of visibility on the steps to know their place*" (S1). They also found that "*the process was long*" (S3). They explained that this was "*linked to recurrent back and forth on elements that were a priori known to all (e.g. the learning outcomes)*" (S3).

The absence of a clearly identified leader from the outset of the process led to misunderstandings and difficulties in managing the process and decision-making. "*A definition of roles and responsibilities not precise enough, project management not rigorous enough and use of human *resources not optimal"* (S6). *"lack of orchestration and follow-up from one session to the next"* (S3). It seemed appropriate to have a leader according to the stages of the process who is an expert in the latter. *"The collaboration was undoubtedly difficult because there was not a leader of the project who would have made a "collaborative point" at the end of each meeting and encouraged people to fill in the documents in a collaborative way. This collaborative writing seems to me to be an important point for the common understanding of the project. But don't you need several leaders depending on the elements to be built (learning outcomes, game metaphor, research, trace, game development, etc...). Move towards collective management?"* (S3).

We used ten of the eleven instruments originally provided, we did not use the instrument for listing didactic obstacles. We developed two additional instruments: Writing the course teaching scenario, Formulating the learning outcomes. Furthermore, specialists proposed six more instruments to assist the team in the process: "Ordering the learning outcomes"; "Choosing the design of the LG"; "Defining a research problem"; "Drawing up an experimental protocol"; "Defining project indicators" and; "Defining the format of traces". Participants could access them to prepare the workshops. These instruments were regularly reviewed and participants were asked to provide feedback. However, some participants felt that "*the tools were too many and too different*" (S3). Involving specialists before the game design begins would allow for the identification of instruments from their practice. The multitude of instruments and their structural diversity hindered the traceability and documentation of collaborative work. Indeed, only the facilitators were able to recall which instruments were used and completed.

### *Conduct of the research and data*

In the context of DBR, developing a research question with teachers entails raising questions about field issues. It was challenging to engage teachers on this aspect of the project. Only the researcher with the most significant interest in the question drafted the problematic based on a theoretical model that they employed in their research. *"It appears to me that the (legitimate) questions of teachers revolve around the effectiveness of the game while the questions of researchers can concern other points (for me, for example, [the questions concern] the experience of the player because the work is part of a more global research program). It is therefore difficult to ensure that the points of view meet."* (S6). For other researchers, the expense of delving into a topic outside of their expertise was too high: "*I don't know this field of research, it will be too costly in terms of time to invest in this new field*" (S1); "*I don't have any theoretical contribution to make on the subject*" (S2). *Furthermore, the work implies the adhesion of the various researchers to a common research paradigm (even if it is multidisciplinary). This is not the case for this project"* (S6). *"a research manager is needed"* (S3).

In the LG design process, teachers' primary objective is to acquire a game that they can use in their class. They may also face time constraints with regards to when they need the game (*e.g.* the game must be ready to be played at a specific point in their lessons). In contrast, research has fewer time constraints. Consequently, LG design by teachers and research linked to a specific problem occur over different timelines. Moreover, they are occasionally motivated by disparate interests.

When a user interacts with the LG, such as a student playing a game, digital traces are produced that can be used to address research questions or meet the needs of teachers. During the project, there were discussions about the content and format of these digital traces as well as the feasibility of analyzing them. "*For the generation of traces, communication was difficult between researchers and developers. To set up the tracing of a game, it was necessary to plan a stage of technical feasibility and verification of the nature of the variables to be generated in the traces. It fails to provide developers with a guide that explains the basis of an actionable dataset, for example,* *showing examples of data on which, it is possible to perform analyses"* (S2). It was necessary to check the consistency and completeness of the digital traces (Berti-Equille, 2012). This work highlighted the role of the data scientist. *"This step requires collaboration between the game developer, a data analyst and a computer scientist capable of taking the traces and formatting them for stats data analysis or identifying algorithms to do so"* (S3).

***Unshared knowledge***

Regarding the sharing of knowledge, one of the first pitfalls concerned the learning outcomes; the participants did not understand each other. The level of granularity was not the same between them. This was an unexpected result. During the eighteenth workshop, a shared understanding of the learning outcomes for LG implementation emerged. This understanding was based on identifying "didactic obstacles" (i.e educational objectives that are difficult to meet).

In this regard, some participants noted the difficulty of aligning educational objectives with the game's principle and how it would be applied. "*The most difficult thing seems to me to be to make the classic disciplinary objectives coincide for the target audience with the principle of the scripted game: not all the notions clearly lend themselves to a transposition in this framework*" (S7).

One other unshared knowledge is the metaphor. One of the researchers suggested the use of a playful metaphor to represent the content to be taught. However, the majority of the team members did not comprehend this proposal and integrating playful and pedagogical elements became a pitfall. This is a well-known difficulty documented in the literature. Furthermore, there were a disagreement regarding the responsibilities for this integration. The teachers believed it was the responsibility of the game designer, while the researchers argued that teachers must achieve this integration as they are specialists in the subject being taught. As explained by participants, the metaphor is quite complex: *"Need to dedicate significant time to understanding and sharing concepts (e.g. metaphor)"* (S3). *"I had a hard time understanding what the game metaphor was"* (S3). *"I think it's a concept that is not shared and therefore not really taken into account in the team"* (S6). *"Find a coherent scenario that allows the targeted notions to come to life in the form of metaphors that can be adhered to without betraying the underlying concepts too much."* (S7). *"Find a metaphor because it was a question of mastering the meaning of the intended educational objective while imagining a way to metaphorize it and to be able to develop it as easily as possible behind it"* (S1). Due to these difficulties, the LG built does not fully comprise a real metaphor.

Similar difficulties were encountered during the research process for the LG. A discussion arose between two collaborating researchers on how to write a specific problem statement using terms like "Analysis model" and "Related work". It took them some time to reach an agreement on this point. Additionally, the research work required adherence to various epistemological paradigms present in the project which was particularly crucial in the context of multidisciplinary research.

In summary, while the prototype was designed based on literature recommendations, existing design models, and inputs from five specialists in the team, persistent difficulties were observed. These difficulties pertain to the complexity of the process, diversity of practices in the field and lack of consensus on its modalities. They also relate to the actors' diversity of expectations and professional cultures, which constitute obstacles to collaborative work, as well as their lack of knowledge of the involved process. As a result, understanding each other and sharing practices even before the start of the design process seems to be an essential prerequisite for effectively carrying out this complex work. Since the process is collaborative, it is crucial to define each person's specialities, roles, and interest. It is also necessary to provide explicit tools for integrating the pedagogical and playful dimensions.

**Discussion: eight principles for LG design to enhance collaboration**

During our workshops, we identified factors limiting the progress of the process, often related to management skills and organization. In addition, we propose eight principles to enhance the collaboration fundamental principle of DBR. They are based on the literature on LG design methods and the results of our study. These principles have the potential to produce specific effects. For some of these principles, we suggest "instruments" to support the design work. To clarify our proposal, we distinguish 3 types of instruments: (1) Educational resources to support the educational component in the institutional context in which the LG is located (*e.g.* the description of the institution's learning outcomes). (2) Guideline that accompany the design stages which include a set of questions to help participants question and build progressively (3) Deliverables that document the design process.

We propose 8 principles:

- N°1- Identify and possess the *necessary skills* for LG design
- N°2- Identify *knowledge that is not shared* to facilitate communication among specialists
- N°3- Co-construct a *framework document* to guide the project, including objectives, instruments, constraints, and participants
- N°4- Co-construct the *design process* to facilitate shared project management
- N°5- Identify the *technical tools* needed for the project and constraints linked to technical developments
- N°6- *Assist and track the process* using relevant and useful tools
- N°7- Facilitate *collaborative research* among LG design specialists using digital traces and produce indicators for evaluating LG
- N°8 - Support the *co-construction of game metaphor* related to learning outcomes

*Principle 1 - Identify and possess the necessary skills for LG design*

Based on our observations during workshops, we have identified several skills that are essential for LG design in a research context, including:

1) Knowledge of the LG design process
2) Familiarity with the subject matter to be taught
3) Understanding the didactics of the subject matter
4) Proficiency in game design, including technical, artistic and playful skills
5) Ability to combine playful and pedagogical elements to develop a game metaphor
6) Mastery of computer development techniques
7) Development of an educational scenario that incorporates the game
8) Knowledge of research methods such as problematization, indicators definition, etc
9) Expertise in data production and analysis methods

We assume that two key skills are necessary:

**Defining and structuring data:** this skill involves ensuring that the data generated by the user's interaction with the LG are accurate and relevant for evaluating the game and answering research questions.

**Articulating the game and pedagogical elements to create a game metaphor:** this skill involves translating the concepts to be taught into a game metaphor that enable the learner/player to acquire targeted knowledge through game interactions.

It is important to note that specialists may have multiple skills.

Expected outcomes of this principle are to possess the necessary skills to design a LG effectively from the beginning.

While there are many recommendations on the need for a multidisciplinary team, the necessary skills are not sufficiently refined, and it would be beneficial to map the skills, knowledge of specialists (principle 2 below) and their interdependencies to improve LG design.

*Principle 2 - Identify knowledge that is not shared*

While principle 1 enables the identification of necessary skills and specialists for the process, it is equally important to identify essential knowledge for the process (e.g. teacher provides all the information on the context, and its pedagogy; game designer provides information on the design of the game). Thus, principle 2 emphasizes the need to identify specific knowledge to lead the design process and facilitate its sharing. The knowledge required for the process must circulate among the involved specialists, so that they can collaborate more effectively. If the knowledge necessary for the process to work must be identified from the beginning, sharing this knowledge must take place at the appropriate point in the process. If knowledge is shared too early, it risks being out of context and therefore useless to non-specialist participants. The knowledge is therefore present latently, i.e. it can be mobilized at any stage of the process. It is therefore essential to first identify and model the knowledge useful for progressing the design process.

During our workshops we identified several useful types of knowledge for guiding the design process, such as learning outcomes, metaphors, digital traces structuring, and problematization.

Expected outcome is to enhance mutual comprehension among specialists, thereby preventing any blockages during specific stages and improving their understanding of the LG design process.

To implement this principle, we can use two theoretical frameworks, praxeology (Chevallard, 1997; Sanchez E. et al., 2017) and boundary object (Carlile, 2004; Star & Griesemer, 1989), to model knowledge and its sharing. process. Other theoretical frameworks on knowledge modeling that could be used and compared to the frameworks cited above.

*Principle 3 - Co-construct a framework document to guide the project*

The study highlights the importance of supervising the process to prevent deviations from the objectives and timeline. A preparatory phase, conducted asynchronously by the project proponent, appears necessary. Its purpose is two folds: firstly, to understand the participants' needs regarding the LG (education, research, etc.) and to identify the necessary documents, tools and resources for the design. Secondly, the participants must describe their LG design skills, as LG design inherently requires multidisciplinary expertise (see principle 1). Based on this information, the project leader can identify when they need to assume a more prominent role in the process and when to take over the project steering. This preparation phase also could identify the knowledge gaps that need to be filled to facilitate knowledge sharing (principle 2). Ultimately, this stage aims to identify the design-related skills that are lacking and to find suitable alternatives.

During this phase, it is crucial to identify the following elements:

- Institutional constraints: These should be evaluated to measure the risks and opportunities of the project. For instance, a SWOT matrix or a strategic planning diagram can be developed to identify these constraints.
- Expectations of participants: It is important to find out their participants' motivations, expectations, and the role they wish to play in the project. Their specialties, skills, and availability must also be identified to determine their level of involvement.

- Skills: Any essential skills needed to lead the project should be identified, and alternatives should be explored if these skills are not available.
- Participants' knowledge level: It is necessary to evaluate the participants' knowledge of LG design. This can be done using a questionnaire that measures their level of understanding of the concepts inherent in LG design.
- Technical: The technical tools required for the LG development process must be considered. For example, if the development platform cannot track the students' activity, an alternative must be identified at the beginning.
- Educational resources: The availability of educational resources, guidelines or other documents to support the design process must be evaluated.

The project leader may distribute a questionnaire to the participants to identify the elements mentioned earlier. Based on the questionnaire's findings, a "framework document" can be drawn up and shared with the participants for their review and validation. Any difficulties and constraints can be discussed to avoid any perceived pitfalls. This framework document is co-written by all the participants, ensuring everyone's input is considered.

Expected outcome is the framing of the project's directions. The team can thus share objectives, motivations, areas of expertise, instruments, etc. The framework document is discussed collaboratively. This document can serve as a reference point for project management and can be refined during the process.

Looking ahead, theoretical models could be employed to establish a guide for designing LG in research settings. Integrating management science concepts presents a promising avenue to explore.

*Principle 4 - Co-construct the design process to facilitate shared project management*

The design process could be co-constructed so that everyone can identify their role and level of involvement. The team should include a specialist in the LG design process who is responsible for outlining the process and explaining each stage to the other participants. This co-construction will enable novices in LG design to be trained before the project is launched. During this collaborative work, each individual well positions itself as a specialist in one or more stages of the process. The process must be able to evolve according to the progress made, any obstacle encountered, or if new participants join the project.

Also, this principle challenges the traditional notion of singular leadership in the design process, and instead promotes shared leadership based on each team member's speciality and the stage of the design process. We propose referring to these leaders as "stage leader", they are responsible for a specific stage of the process and have the final decisions-making authority and accountability for summarizing the stage's outcomes. They must ensure that all necessary documents are prepared for traceability decision-making throughout the process. It seems crucial to identify each stage leader during the co-construction of the process.

Expected outcome is to establish a co-constructed process that is familiar to all participants. This process will involve identifying specialists and leaders to 1) manage the steps, 2) ensure understanding between the participants, 3) write the stage deliverables and 4) ensure reliable decision-making.

In order to implement this principle, a flexible process can be created using a process design language. We have begun this work on using the THEDRE language (Mandran N. & al., 2017) and we plan to test it through a game during a workshop in June 2023.

***Principle 5 - Identify the technical tools and constraints due to technical developments***

The selection of a technical platform for developing the LG seems a crucial design stage that require the expertise of game development specialists. This decision determines content and format of the digital traces generated, and it should be made with three key stakeholders: 1) game development specialists, who are knowledgeable about the traces that can be produced by the platform; 2) data production methods specialist, who can ensure the coherence and completeness of the data for analysis purposes; and 3) researcher, who requires the traces to construct indicators that address their research questions.

Expected outcome is to provide a technical description of the traces generated by the LG, including a data dictionary (*i.e.* a dataset description), metadata, and other relevant information. This deliverable enhances communication among IT developers, data analysts and researchers.

In terms of perspectives, a tailored model for LG could be developed using established standards like xAPI that are geared towards tracking and recording data.

***Principle 6 - Assist and track the process using relevant and useful tools***

The multidisciplinary LG design process is time consuming and requires proper documentation to ensure traceability of the process. This is particularly crucial since the process involves research stages where scientific findings must be documented. Educational resources and guidelines can help the stage leader in this regard. It is necessary to identify the resources commonly used by the participants, which are aligned with their practices and knowledge. Guidelines comprise a set of questions designed to simulate reflection (Mandran N. et al., 2022). This is a means of transferring expertise because each question refers to what stage leaders themselves ask to conduct their work.

The stage leader must establish the guidelines based on this knowledge to meet the expectations of each of the design stage. The stage leader facilitates the session with the set of questions, writes a synthesis and produces the stage deliverable (if required, outside the workshop). Then, each deliverable is shared and validated by the participants and these deliverables are used for decision-making. The guidelines for each significant stage of the process could include: "Defining the project", "Choosing the learning outcomes", "Building a research problem and indicators", "Scripting the LG".

Expected outcome is to enhance communication among participants by providing dedicated guidelines for each stage. In each stage, the stage leaders share their expertise by asking questions.

Some guidelines created during our workshops need to be tested in other contexts and by different users to evaluate and improve them. Additionally, new guidelines should be developed, particularly to support the construction of the metaphor. It will also be interesting to test the transfer of knowledge between participants on a larger scale based on these guidelines.

***Principle 7 - Facilitating collaborative research among LG design actors using digital trace and producing indicators for LG evaluation***

Our research aims to address questions from teachers who will use a LG in their course. We aim to involve all participants, including the teachers, in the problem construction. To accomplish this, we have mobilized five specialists: the teacher, the game designer, the researcher, the data analyst and the technical developer. To support this principle, we propose a model (Fig3) that highlights the specialists, the useful knowledge to facilitate the collaboration and relation between specialists.

Collaboration between specialists is built around this knowledge. Nevertheless, during our workshops we were able to note the difficulty of working together at all stages. It therefore seems necessary to us not to share all knowledge among all. Some specialists will collaborate to develop a deliverable (in the sense of a shareable document) which will then be shared with the other specialists involved in the rest of the process.

To illustrate our point of view, we describe some relation between specialists. The teacher designs an educational scenario (step 1) that he shares with the game designer to design LG (step 2). The teacher also shares a field question (step 3.1) so that the researcher develops a problematic based on theoretical models (step 4). For the collection of digital traces, the data analyst develops a set of measures and variables to address the research questions in collaboration with the researchers (step 5), along with a set of quality criteria to qualify the data (Di Ruocco et al., 2012). To do this, they rely on the data dictionary provided by the developer (step 3.2). The developer then provides a qualified and analysable file to the analyst (step 6). The analyst checks the data provided and may return to the developer to improve the content of the file. For the collection of digital traces, IT developer collaborates with game designer to identify which data supplied by LG. After the analysis (step 7), the analyst reports the results and indicators to the researcher with results and indicators. Upon reviewing the results, the researcher may request additional analyses, indicators or even other data. The researcher interprets the results (step 8) and provides pragmatic solutions to answer the teacher's questions in the field. (step 9). Teachers and researchers discuss the results and the solution in order to objectify them with the points of view of the teachers and the fieldwork constraints (step 10) and game designer to enhance the LG (step11).

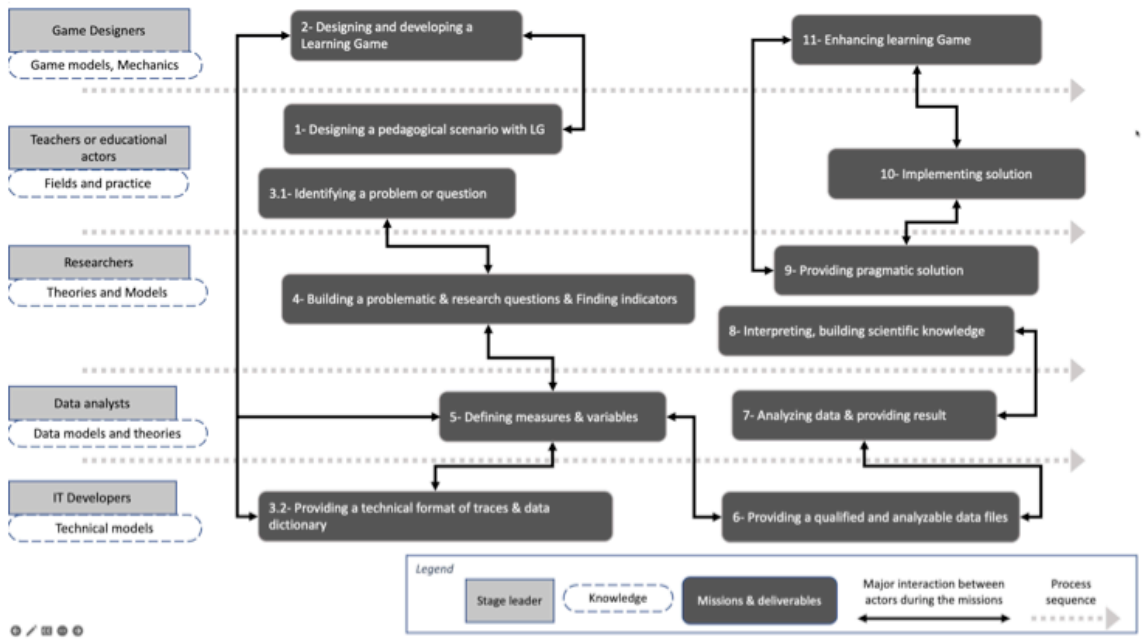

Figure 3 - Specialists, knowledge and main interactions .

Expected outcomes are (1) to build a problematic based on teachers' questions and (2) to ensure the production of relevant indicators in collaboration with data analysis specialists.

The proposed process is still in its initial stages and can be improved; especially in the creation of indicators to answer research questions. The role of the data analyst appears crucial in managing the link between the researcher and developer to propose an adaptable method. Additionally, the co-problematization stage between researcher and teacher needs to be clarified and modelled. It is also worth considering the specificity of collaborative research on LG.

***Principle 8 - Support the co-construction of the game metaphor related to the learning outcomes***

As stated by (Fabricatore, 2000), intrinsic metaphors are highly effective as they are part of the LG. The quality and efficiency of a LG depends on the integration of game and learning components through the metaphorization of the subject to be taught (Bonnat et al., 2023) and the deconstruction of this metaphor during the debriefing phase after the game's use (Plumettaz-Sieber, Bonnat, et al., 2019). However, designing a coherent and implicit metaphor that aligns with the LG's universe and context can be challenging for non-specialists. Nonetheless, it is crucial to ensure that the user can effectively apply the learning outcomes during the game session (Bonnat et al., 2023).

Expected outcome is the integration of a relevant metaphor in the LG that have the potential to foster learning.

The relationships between learning outcomes, metaphor, game world and game mechanics is seldom studied despite being crucial to effective learning through LG. To address questions about their relevance for learning, it is necessary to give full attention to this aspect and explore it in detail by proposing a model and specific working methods.

It is important to note that the 8 principles are interconnected. However, some of them such as N°1 Skill, N°2 Knowledge, N°3 Framework Document, and N°4 Co-Design Process, must be initiated before entering the design process while principle N°5, about technical choices, must be considered at the start of the process and not revisited later to avoid time costs. Other principles, like N°2 Knowledge and N°6 Assist and Trace, are cross-cutting principles that support project monitoring.

**Conclusion**

Our paper presents a study on pitfalls and recommendations from the literature and their implementation in real context.

This study enabled us to test LG design models in a collaborative research setting using a prototype to design a new LG. Despite incorporating the different models, the use of the prototype revealed the same pitfalls: 1) complexity of the process; 2) communication among the multiple specialists involved in the co-design process; 3) difficulty of balancing and articulating the playful and educational elements in a coherent manner; 4) sharing knowledge among different specialists, and the lack of research conducted in collaboration with domain specialists.

We used the method of participatory observation over an extended period and were actively involved in the study. This method of generating contextual data is logistically challenging to set up and demanding in terms of participant involvement. However, it allowed for a thorough examination of real difficulties.

This project helped identify key aspects of game design and the need to prepare for and support this complex process. To this end, we proposed 8 principles to support the LG co-design process: (1) Identify and possesses the necessary skills for LG design; (2) Identify unshared knowledge and expertise; (3) Co-construct a framework document to guide the project; (4) Co-construct the design process for shared project management; (5) Identify the technical tools needed for the project and constraints linked to technical developments; (6) Assist and track the process with relevant and useful tools; (7) Facilitate collaborative research among LG design stakeholders based on digital trace and produce indicators for LG evaluation; (8) Support the co-construction of the game metaphor.

Although many works exist in the literature, they are most often conducted on the overall process. The proposed principles make it possible to specify the major points of difficulty in the design of LG and to work on specific points of this process. Additionally, they can help to clarify questions and issues surrounding LG designs. Consequently, research questions could be conducted for each of these principles.

Adhering to these principles would enhance the efficiency, cost-effectiveness and usability of LG's design. Collaborative efforts would also enable the exchange of concepts and expertise among education professional working game-based pedagogy.